\documentclass{aa}
\input psfig.sty
\begin{document}
\title{Radio Spectral Index Variations of the SNR HB3}
\author{Wenwu Tian\inst{1,2}
      \and
        Denis Leahy\inst{2}}
\offprints{W. W. Tian}
\institute{National Astronomical Observatories, CAS, Beijing 100012, China; tww@ns.bao.ac.cn\\
\and Department of Physics \& Astronomy, University of Calgary, Calgary, Alberta T2N 1N4, Canada}

\date{Received December 1, 2004; accepted on Jan. 26, 2005} 

\abstract{We present images of the SNR HB3 at both 408 MHz and 1420 MHz from the
 Canadian Galactic Plane Survey mainly based on data from the Synthesis Telescope 
of the Dominion Radio Astrophysical Observatory. We reproduce the 100m-Effelsberg
 HB3 image at 2695 MHz at large scale, and find that there exists a background
 emission gradient across the HB3 area. Based on our analysis of background emission 
and the boundary between W3 and HB3, we 
give HB3's flux density as 68.6$\pm$11.5 Jy at 408 MHz and 44.8$\pm$12.0 Jy 
at 1420 MHz, after subtracting flux from compact sources within HB3. The 
integrated flux-density-based spectral index between 408 MHz and 1420 MHz is 
0.34$\pm$0.15. The averaged T-T plot
 spectral index using all subareas is 0.36. Our measurement values are less than a
 previously published value of 0.6. The 408-1420 MHz spectral index varies spatially
 in HB3 in the range 0.1 to 0.7. We investigate the data used by previous authors, 
and consider 
more data at 232 MHz, 3650 MHz and 3900 MHz which are not included in previous
 calculations. There is evidence for two spectral indices for HB3 in 
the radio band, i.e., 0.63 (38 - 610 MHz) and 0.32 (408 - 3900 MHz). This is 
consistent with the spatial variations: the low frequency data mainly reflects the 
steeper indices and the high frequency data mainly reflects the flatter indices. 

}
\titlerunning{Radio Spectral Index of HB3}
\maketitle

\section{Introduction}
   The evolved supernova remnant (SNR) HB3 (G132.7+1.3; Landecker et al., 1987), as 
one of the largest galactic SNRs currently known, has been previously observed by 
Caswell J.L. (1967) and 
Landecker et al.(1987) in the radio continuum, by Leahy et al. (1985) in X-rays, by Routledge 
et al. (1991) in the HI-line, by Fesen et al. (1995) in H$\alpha$, and by Koralesky et al. 
(1998) in hydroxy1 (OH). Its known basic physical features are: 60$\times$ 80 pc in 
size based on a distance of 2 kpc (Routledge D, 1991). It interacts with 
the molecular gas in which W3 
is embedded, but there is no evidence for a direct interaction of W3 and HB3; strong radio-optical 
correlation is also observed. There is no detailed study on HB3's radio
 spectrum so far. The 
complex surrounding of the SNR embedded in a star-forming region that contains the 
W3 HII complex makes it difficult for previous authors: to spatially 
separate HB3 from the W3 complex, and to estimate the background emission level in 
order to obtain an accurate integrated flux density of HB3. 
Many previous radio observational data on HB3 are of limited use due to the 
limited spatial resolution. Although a spectrum index for HB3 of about 0.6 has been 
suggested, it is worth further study.  Therefore, we use new HB3 data from the 
recently finished Canadian Galactic Plane Survey, described in section 2, and present 
the results and a discussion of HB3's spectrum in section 3. 
A summary of findings is given in section 4. 

\section{Observations and Image Analysis}
     The 408 MHz and 1420 MHz data sets come from the Canadian Galactic Plane 
Survey (CGPS) which is described in detail by Taylor et al. (2003). The radio 
continuum 
observations' resolution is below 1'$\times$ 1' cosec($\delta$) at 1420 MHz and 
3.4'$\times$3.4' cosec($\delta$) at 408 MHz, which is the highest available 
resolution for HB3 so far. Because  
Dominion Radio Astrophysical Observatory (DRAO) Synthesis Telescope (ST)
observations in the CGPS are generally 
not sensitive to structures larger than an angular size scale of about 3.3$^{o}$ at 
408 MH and 56' at 1420 MHz, the CGPS includes data from the 408 MHz all-sky survey 
of Haslam et al. (1982) which has an effective resolution of 51' and the Effelsberg 
1.4 GHz Galactic plane survey of Reich et al. (1990, 1997) with resolution 9.4' for 
large scale emission (the single-dish data are freely available by http://www.mpifr-
bonn.mpg.de/survey.html). 
Taylor et al. (2003) mentioned artifacts of the CGPS images arising from the single-
antenna data. We check the HB3 map from the 408 MHz all-sky survey, and confirm that 
there exists a type of artifact, i.e. low-level striping that appears as 
discontinuities across lines of constant Right Ascension. The maximum amplitude of 
the striping with scale of 3$^{0}$ is less than 6$\%$. In the 408 MHz single-antenna 
map, HB3 is located between two adjacent strips, and is very little contaminated by 
the artifact, i.e, only part of B1 region of HB3 (see Fig. 1)  is influenced by the 
artifact. In the 1420 MHz Effelsberg map of the HB3, sidelobes have positions to the 
east and west from the origin. However the amplitude of the effect is less than 
2$\%$ of the brightness of HB3 so it is not very significant.
Since the CGPS integrates both single-antenna  and interferometer data, information 
on HB3's structure  on all angular scales larger than the resolution 
limit is included.  By using a special 
image-processing procedure, the dynamic range of the continuum images reaches up to 
10000 at 1420 MHz and up to 5000 at 408 MHz, yielding essentially noise-limited 
images with an rms of ~0.3 mJy beam$^{-1}$ at 1420 MHz and ~3 mJy beam$^{-1}$ at 
408 MHz (Taylor et al., 2003).

   Estimating HB3's flux density at each frequency is the first important step to 
study its spectrum variation. We analyze the HB3 images using the  DRAO export
software package to get HB3's flux density at each frequency. 
To study thermal contamination from W3 and keep it as small as possible, 
we study the radio spectrum in 16 different regions in HB3. 
The influence of compact sources within HB3 is very much 
reduced by two alternate methods: 
a) compact sources within HB3 are each fit by Gaussian components plus 
a twisted plane background, then the source component is removed from the image; 
b) a small area including each compact source is completely removed from the
 analysis. Our analysis shows that the latter method reduces the effect 
of compact sources on the spectral index of HB3 to minimum. 
 
\section{Results and Discussions}
\subsection{HB3 Structure at 408 MHz and 1420 MHz}
  The CGPS maps at 408 MHz and 1420 MHz are shown in the upper left and right 
panels of Fig. 1.   
The lower left panel shows the 1420 MHz map which has been convolved to the same 
resolution as the 408 MHz map prior to source removal. For reference we also 
reproduce the 2695 MHz Effelsberg map (lower right; F\"urst et al., 1990) in Fig. 1. 
The Effelsberg map has an resolution of 4.3' and a sensitivity of 50 mKT$_{B}$. 
The 408 MHz CGPS image is generally similar to previously published 408 MHz 
images (Landecker et al., 1987 and Fesen et al., 1995), but shows significantly
more detail and clearer filament features, as well as greatly reduced artifacts due
to compact sources and W3. This is due to the higher sensitivity and dynamic range 
of the CGPS data.  
Our 1420 MHz map shows rich weak emission features with much better quality 
than the Fesen et al. (1995) map. The CGPS maps at both frequencies are free from 
any strong artifacts caused by W3 or due to strong compact sources, which appear in
the Fesen et al. maps. 
The compact sources within HB3 are well resolved at both frequencies. The higher 
spatial resolution at 1420 MHz means more compact sources are seen. 
Both 1420 MHz and 408 MHz CGPS maps have sufficient spatial resolution to identify
and correct for all compact sources that have significant flux.

\begin{figure*}
\vspace{90mm}
\begin{picture}(200,300)
\put(-18,225){\includegraphics{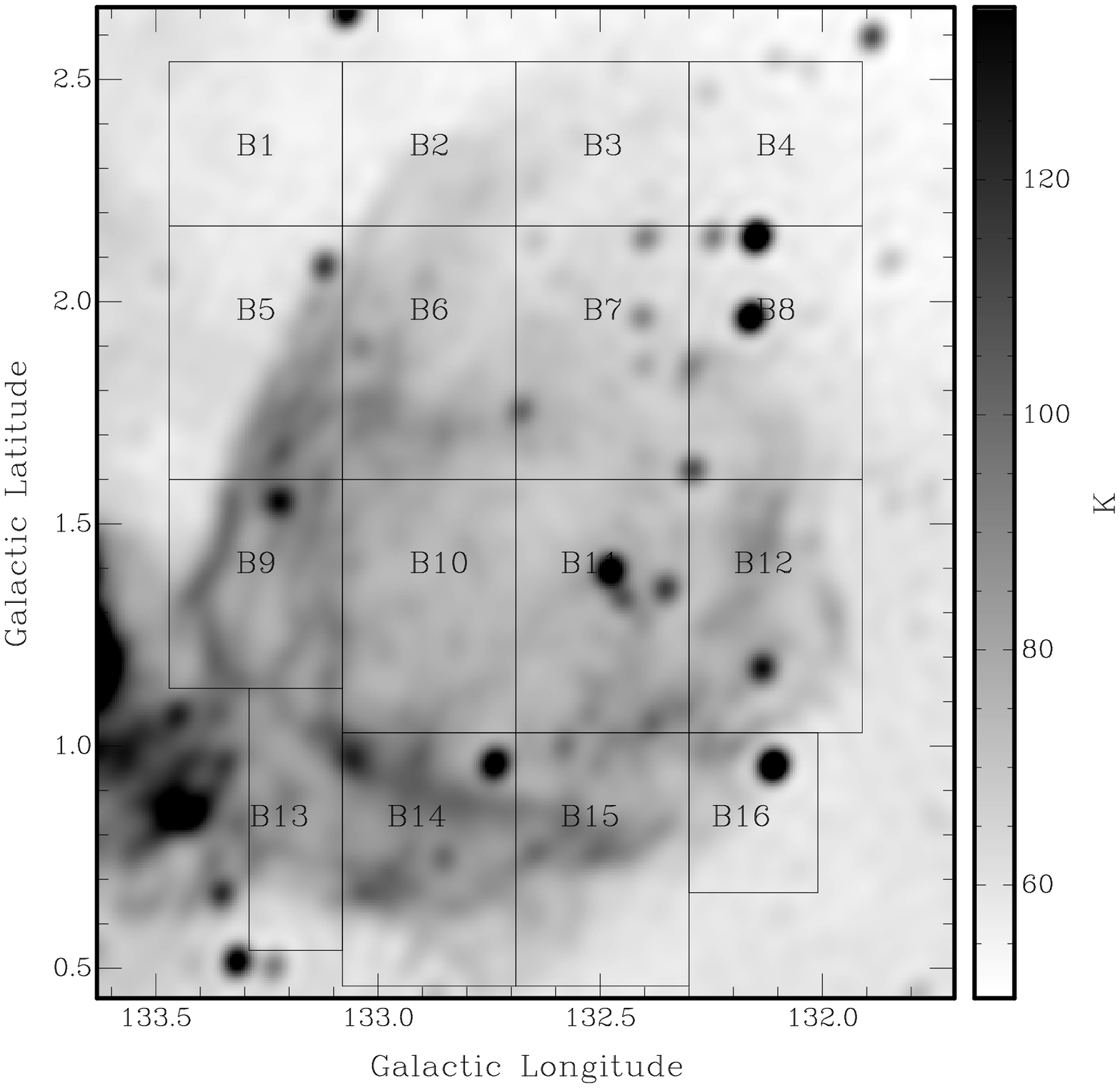}}
\put(256,225){\includegraphics{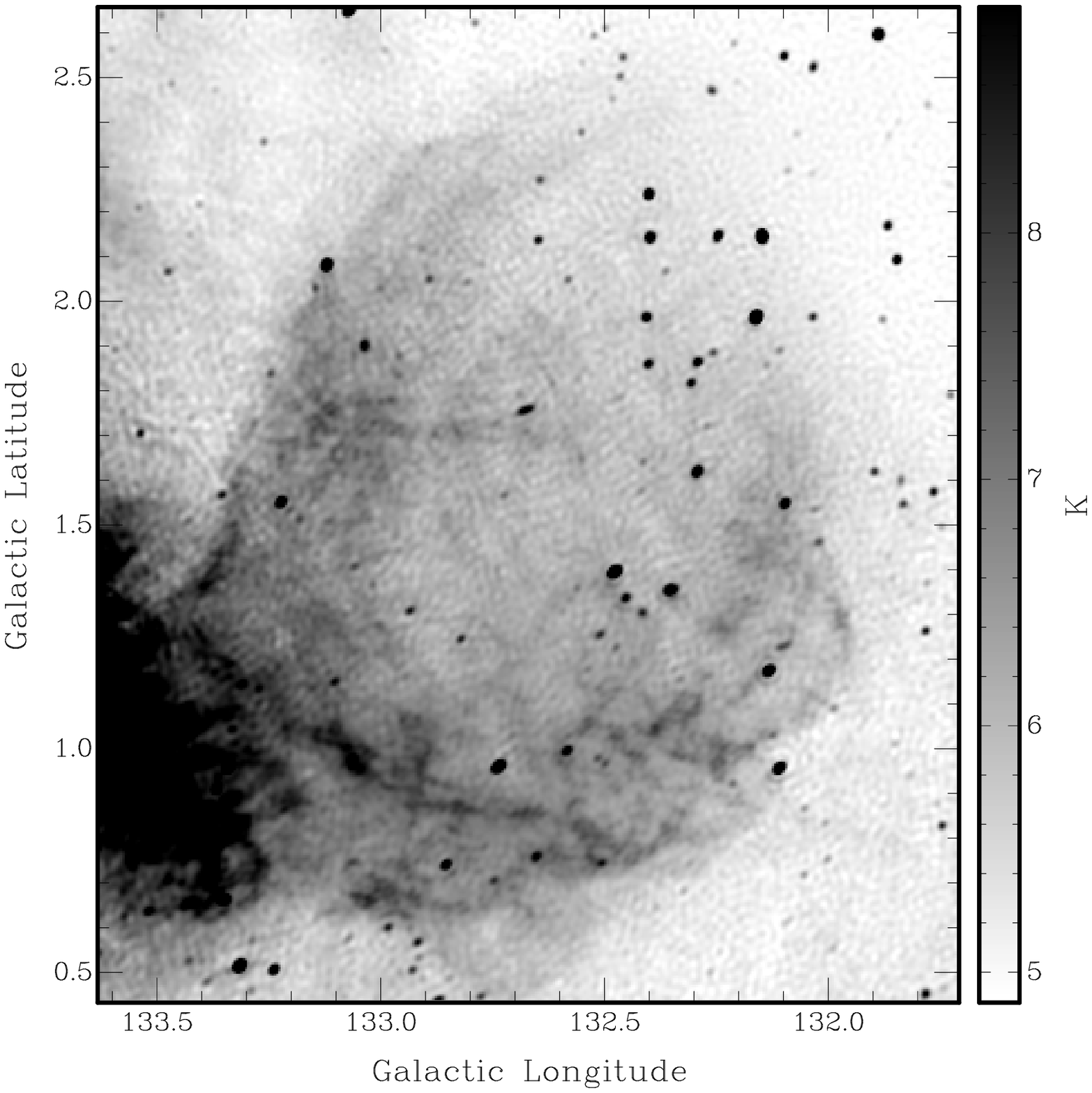}}
\put(-18,-50){\includegraphics{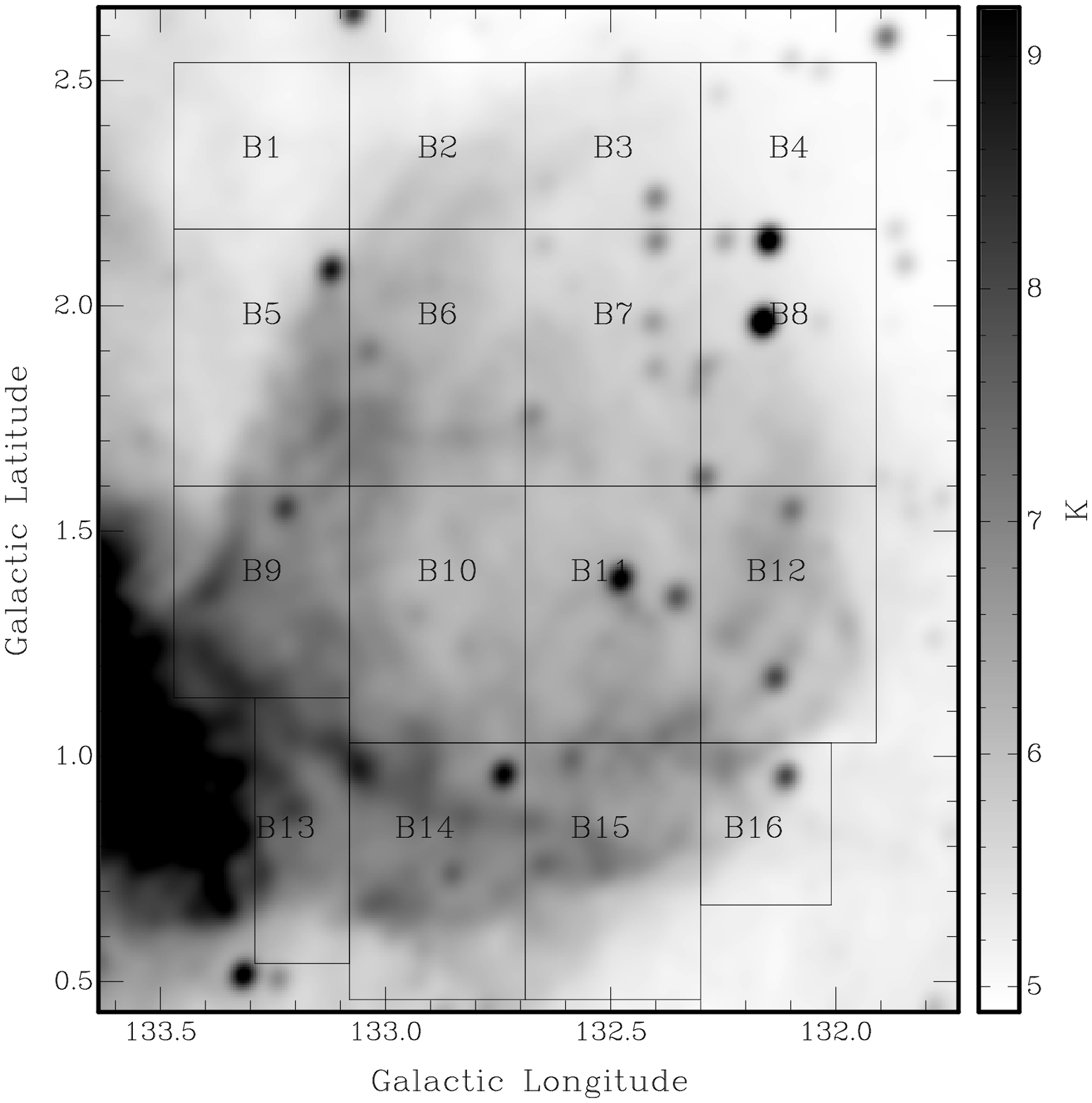}}
\put(190,270){\includegraphics{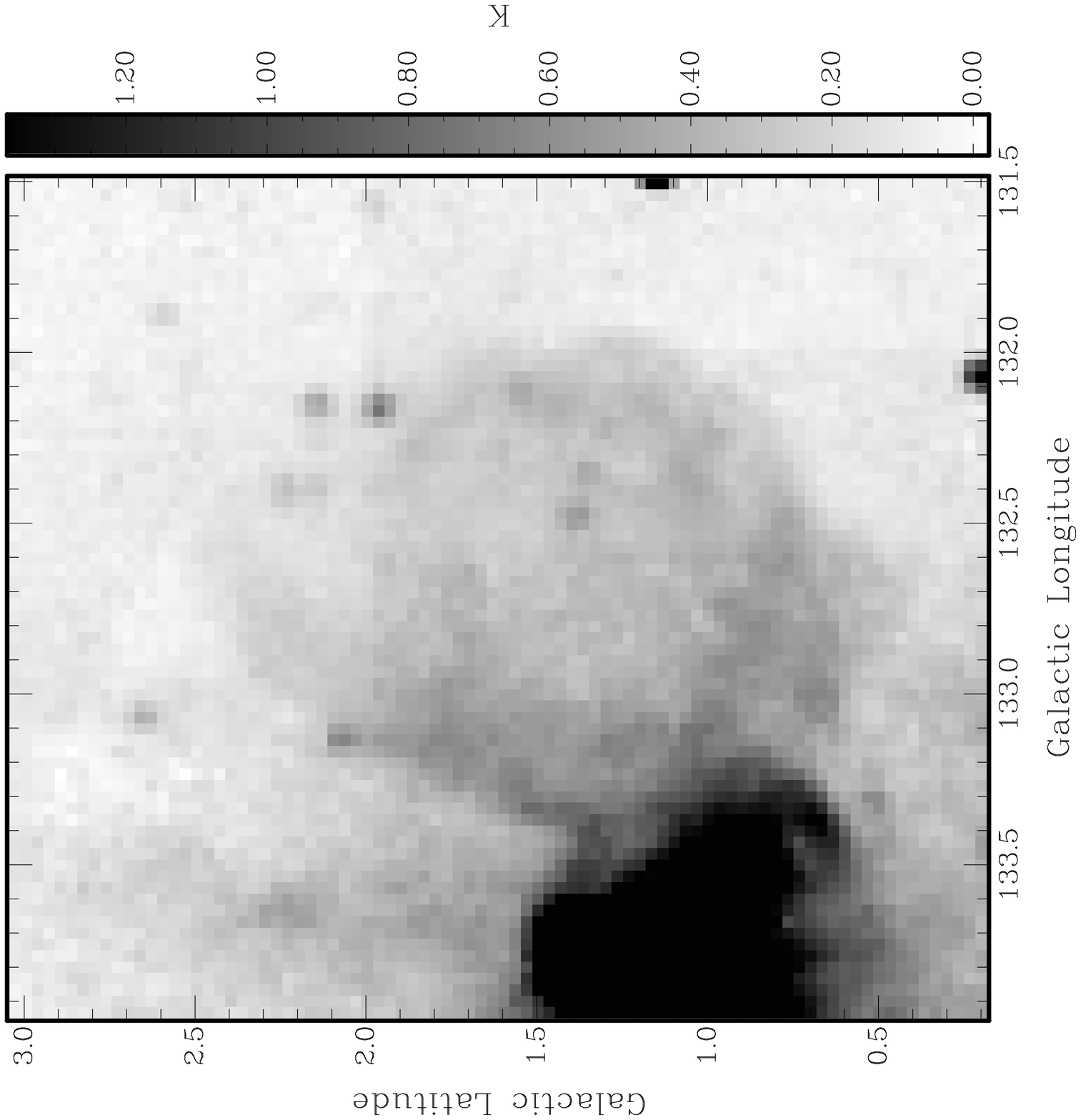}}
\end{picture}
\caption[xx]{The first row shows the HB3 CGPS maps at 408 MHz (left) and 1420 MHz
 (right).  The left panel of the second row shows the 1420 MHz map convolved to the
 same resolution as the 408 MHz map. The 2695 MHz Bonn map is shown at right in the
 second row. The 16 boxes used for T-T plots are shown in the upper and lower left
 panels.}
\end{figure*}

\begin{figure*}
\vspace{30mm}
\begin{picture}(60,100)
\put(-55,190){\includegraphics{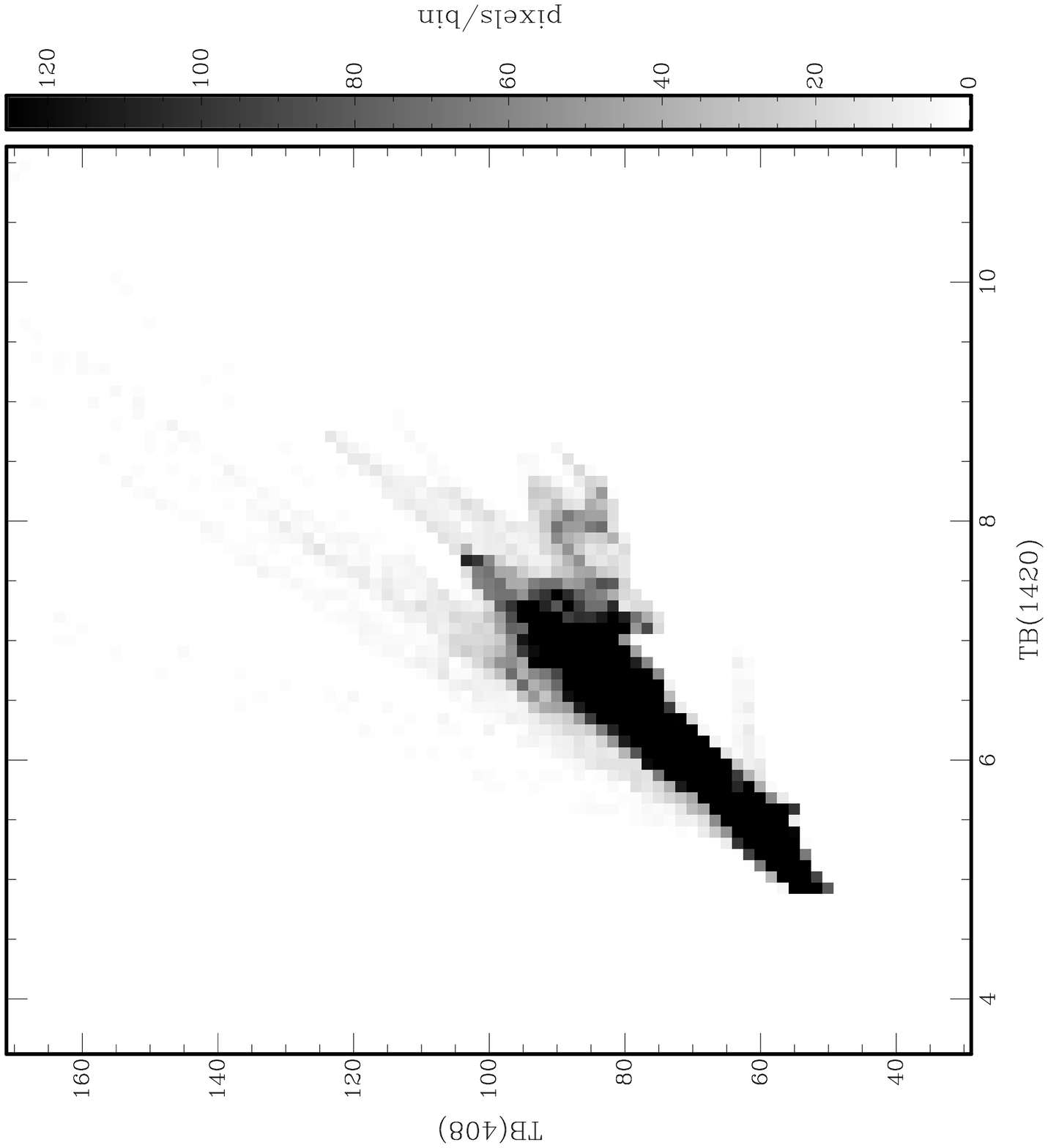}}
\put(118,190){\includegraphics{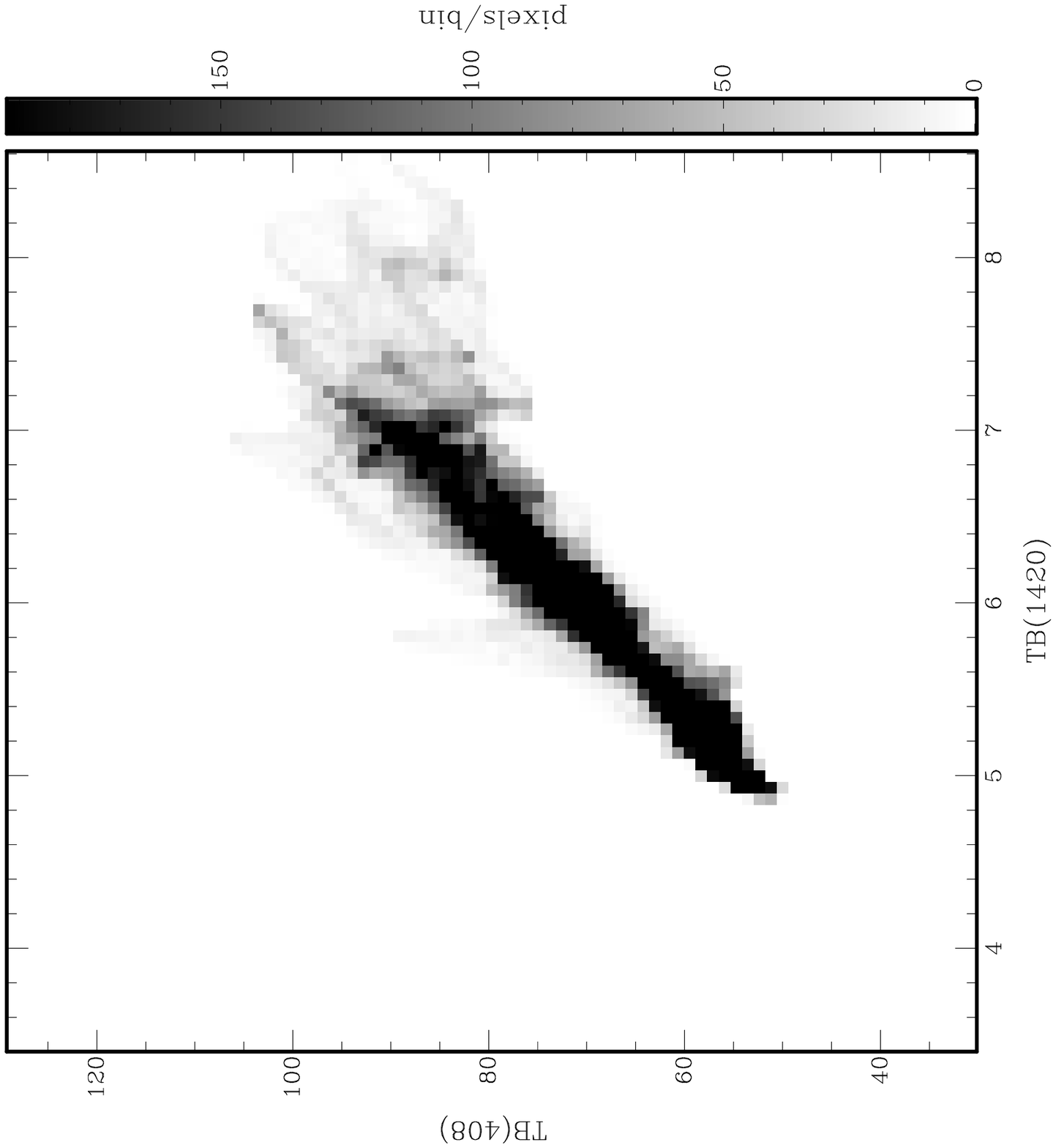}}
\put(285,190){\includegraphics{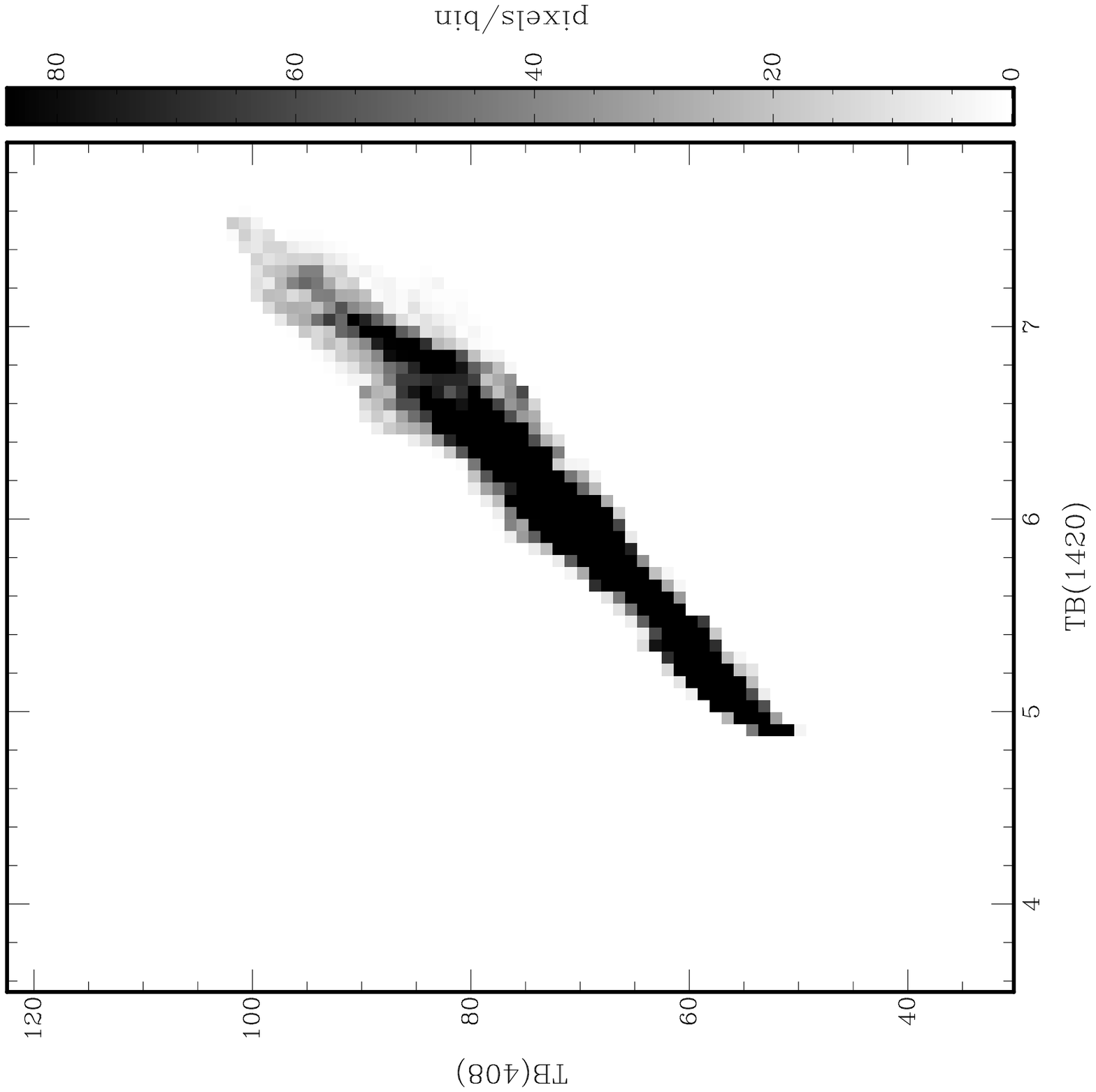}}
\end{picture}
\caption[xx]{408 MHz - 1420 MHz T-T plots using one box covering HB3, 
from left to right: plot for map including compact sources ($\alpha$=0.25$\pm$0.12);
 plot for map with Gaussian fits to compact sources subtracted 
($\alpha$=0.19$\pm$0.11); plot for regions including compact sources removed 
from analysis ($\alpha$=0.22$\pm$0.06).}
\end{figure*}

\subsection{HB3 Spectral Index}    
  HB3's 408-1420 MHz spectral index $\alpha$ (S$_{\nu}$=S$_{o}$$\nu$$^{-\alpha}$) 
is either about 0.34  based on our direct calculations by using integrated flux
 densities at both frequencies or 0.36 by the T-T plot method.  The principle of
the T-T plot method is that brightness temperature spectral indices 
(T$_{\nu}$=T$_{o}$$\nu$$^{-\beta}$) are calculated by fitting a linear relation 
to the T$_{1}$-T$_{2}$ values of all pixels within a given map region. Here 
T$_{1}$ is the brightness temperature of a given map pixel at the first frequency. 
T$_{2}$ is the brightness temperature of the same pixel at the second frequency. 
The brightness temperature spectral index $\beta$ is then derived from the slope 
of the curve. The flux density spectral index $\alpha$ is related to the
 brightness temperature spectral index $\beta$ by $\beta$=$\alpha$+2. We use
 spectral index to refer to the flux density spectral index $\alpha$ in this paper
 unless we specifically say otherwise. The T-T plot method has widely been accepted
 and used for spectral index calculation (e.g., Zhang et al., 1997; 
Leahy and Roger, 1998). 

In Leahy and Roger (1998), the effect of compact sources is reduced by avoiding 
including any strong sources in the regions analyzed.  
For the T-T plot spectral index, we choose a single analysis box covering
131.935$^\circ$ to 133.285$^\circ$ Galactic longitude, and 0.725$^\circ$ to 
2.705$^\circ$ Galactic latitude. This box is conservatively chosen to include most
of HB3 yet have its leftmost boundary clearly right of the edge of W3.
This region yields the T-T plots shown in Fig. 2 and the spectral index 
for the three cases: including, subtracting and excluding compact sources 
within HB3 (see Fig. 2 caption for details). 
Since the brightest compact sources have a steeper spectrum than HB3, they show
up in the T-T plot as the steeper lines of points (Fig. 2 left panel).
 The middle panel is the T-T plot from the map with compact sources subtracted by
subtraction of the Gaussian components fit to the original map.
 In both the left and middle panels, there is a significant amount of emission 
to the right of the main emission band from HB3. This emission has an irregular
structure which has no simple interpretation in terms of a spectral index.
 This irregular structure could be due to weak artifacts from the compact 
sources, which are present before and after Gaussian component subtraction. 
The second method we used to remove compact sources is to remove a small region
including the compact source from the analysis. The region is taken to be a few 
beamwidths wide, so that any contribution from the compact source is below 1$\%$ 
of the diffuse SNR emission.  
Thus when a compact source is removed, any artifacts associated with the compact
source are also removed. 
This is proven by the right plot of Fig. 2, where the compact sources have been
removed and the T-T plot does not show the irregular structure.  

\begin{table*}
\begin{center}
\caption{HB3 408-1420 MHz Spectral Index for Regions B1 to B16}
\setlength{\tabcolsep}{1mm}
\begin{tabular}{cccc}
\hline
\hline
 Sp. Index   |    &$\alpha$ & $\alpha$& $\alpha$ \\
\hline
\hline
 Area No. \vline &including compact sources& compact sources subtracted & compact sources removed\\
\hline
 B1& 0.06$\pm$0.50& 0.03$\pm$0.73&0.07$\pm$0.49 
\\
 B2& 0.29$\pm$0.07& 0.29$\pm$0.07&0.29$\pm$0.07 
\\
 B3& 0.12$\pm$0.21& 0.18$\pm$0.23&0.18$\pm$0.14 
\\
 B4& 0.46$\pm$0.09& 0.30$\pm$0.20&0.28$\pm$0.14 
\\
 B5& 0.51$\pm$0.06& 0.54$\pm$0.06&0.55$\pm$0.05
\\
 B6& 0.43$\pm$0.06& 0.46$\pm$0.08&0.45$\pm$0.08 
\\
 B7& 0.52$\pm$0.05& 0.43$\pm$0.07&0.31$\pm$0.09 
\\
 B8& 0.45$\pm$0.04& 0.31$\pm$0.05&0.25$\pm$0.02 
\\
 B9& 0.12$\pm$4.15& 0.05$\pm$4.10&0.11$\pm$0.79 
\\
 B10&0.12$\pm$0.10& 0.09$\pm$0.10& 0.09$\pm$0.10
\\
 B11 & 0.60$\pm$0.12& 0.39$\pm$0.12&0.40$\pm$0.07 
\\
 B12 & 0.45$\pm$0.06& 0.41$\pm$0.07& 0.36$\pm$0.04
\\
 B13 & 0.10$\pm$0.14& 0.07$\pm$0.13&0.05$\pm$0.13 
\\
 B14 & 0.45$\pm$0.02& 0.38$\pm$0.04&0.45$\pm$0.01 
\\
 B15 & 0.29$\pm$0.02& 0.32$\pm$0.03&0.29$\pm$0.02 
\\
 B16 & 0.25$\pm$0.03& 0.25$\pm$0.03&0.23$\pm$0.02 \\
\hline
\hline
\end{tabular}
\end{center}
\end{table*}
In view of a likely gradient of the thermal emission due to W3, we consider the
 effect of such a gradient on the T-T plot spectral index. For a large area such 
as the box for the whole of HB3 the thermal emission component has a significant
 variation which is larger at 1420 MHz than 408 MHz. When this is added to the 
non-thermal emission and a T-T plot is made, the set of points is broadened in 
the direction parallel to the T$_{B}$(1420) axis more than in the direction 
parallel to the T$_{B}$(408) axis. The broadening is equal to the difference 
in brightness temperature due to the thermal component across the map at each
 frequency. If we choose smaller boxes, e.g. 1$/$4 size in linear dimension, the
 broadening due to difference in thermal emission across the box is reduced by
a factor of 4. Thus we consider smaller boxes to reduce the effect of a thermal
 emission gradient. There is a limit on how small one goes, as the box needs to be
 large enough to contain significant variation  in the non-thermal emission. 

Thus we divide HB3 into the 16 areas shown in Fig. 1. 
The areas were chosen to cover HB3 completely, then the regions B9 and B13 
were adjusted to include the leftmost part of HB3 while avoiding emission from 
W3 as best as possible.
For each region the spectral index was obtained between 408 MHz and 1420 MHz by
the T-T plot method. Table 1 lists the results for the three cases of analysis: 
including compact sources, subtracting compact sources, and removing compact 
sources. There is not a large difference in results between the three methods,
 however visual inspection of the T-T plots shows that the third method produces 
the most reliable results, as it did for the large single box covering HB3 
shown in Fig. 2.
We mention that compact sources' influence on the spectral index calculation 
is obvious in some areas such as boxes 4,7,8 and 11. Thus from now on we discuss
spectral indices derived with compact sources removed, unless specified otherwise.
The spectral index of Box 1 is very flat with a large error. As we mentioned in
 section 2,  Box 1 is influenced by the artifact arising from the 408 MHz single
 antenna data, and in fact is mostly outside the HB3 emission area; this is a
 measure of the thermal background emission. The derived spectral index values are
 between 0.2 to 0.5 except those for Boxes 9, 10 and 13. The averaged value of
 spectral index excluding large error regions B1 and B9 is 0.30$\pm$0.07. This is
 larger than the value of the whole of HB3 of 0.22$\pm$0.06 (see caption of Fig.2),
 consistent with a reduced contribution from the thermal gradient.   

Because the T-T plots of boxes 2, 5, 6, 9, 10, 13 show a complex 
distribution indicative
of multiple spectral indices, we divide each of them into four equal-size 
subareas: a) upper left; b) upper right; c) lower left; and d) lower right.
In many cases the subdivision separates areas of different spectral index so that
the T-T plot method can yield a useful spectral index. 
Table 2 gives the spectral index of each subarea obtained by the automated linear
fit to the T-T plot and by a manual linear fit in columns 2 and 3, resp., 
and the average value of the manual linear fits for each box in the final column.
When there are multiple lines of points in the T-T plots, the manual linear fits 
give a more reliable spectral index.
The values in Table 2 reveal that the spectrum index can be different in different
subareas or even vary within a single subarea. 
The average values for B2, B5 and B6 is similar to those given in Table 1, but
for B9, B10 and B13 the values from the more detailed analysis differ significantly
from that from the automatic linear fit. This is due to the error in fitting a
mixture of lines of data points with a single model line.
We note that B9 and B13 are the two boxes closest to W3 and when we fit the subboxes
using manual linear fits, the correct T-T plot spectral indices are 0.45 and 0.33.
These values are consistent with spectral indices for the other regions of HB3.  
The average spectral index
of the whole of HB3 (excluding region B1) using areas and subareas is 0.36. 
This is the best value we can get using the T-T plot method.
 
Next, in order to continue analyzing the thermal background influence on HB3 in
 detail,  we reproduce the 2695 MHz Effelsberg HB3 map in Fig. 1. The map shows
 there generally exists a uniform thermal background emission gradient from 400mK
 nearby the border between W3 and HB3 to about 80mk at HB3's northwestern edge. 
This gradient is also seen at a lower level in the 1420 MHz map. Considering the
 general background emission gradient, we calculate the integrated flux density of
 HB3.  With  background values averaged from around the whole boundary of HB3, the
 integrated flux densities are 68.6$\pm$11.5 Jy at 408 MHz and 44.8$\pm$12.0 Jy at
 1420 MHz (see Table 3). This choice of background subtracts any smooth  gradient of
 background emission from HB3.  Here we have used the compact source removed maps at
 both 408 MHz and 1420 MHz and also corrected for the area of HB3 removed in the
 compact source removal. The background uncertainty and the calibration uncertainty
 both contribute significantly ($\sim10\%$) to the uncertainty in integrated flux.

The resulting spectral index is 0.34$\pm$0.15 between 408 MHz and 1420 MHz.
Table 3 gives fluxes and spectral indices for HB3 as a whole and for five subareas
 with background calculated separately for each subarea. The  subareas have compact
 sources removed (the subarea definition is given
in Table 3). For comparison, fluxes and spectral indices for the whole of
HB3 are given including compact sources as well for compact sources removed,
and the sum of fluxes of compact sources within HB3 is given.
The results show that compact sources contribute less than 7$\%$ to HB3's flux
and only have a small effect on the spectral index.
Also, except for subarea 3, the subareas' $\alpha$ values are consistent
with the spectral index 0.34$\pm$0.15 for the whole of HB3.
Subarea 3 is the region most likely contaminated by additional non-smooth thermal
 emission from W3. 
\begin{table}
\begin{center}
\caption{HB3 408-1420 MHz Spectral Index for Subareas}
\setlength{\tabcolsep}{1mm}
\begin{tabular}{cccc}
\hline
\hline
 Sp. Index  | &$\alpha$&$\alpha$&$\alpha$\\
\hline
\hline
Area No. \vline&automatic fit& manual fit& box average\\
\hline
 SUBBOX 2a& -0.2$\pm$4.0   &0.3 &  0.35
\\ SUBBOX 2b& 0.3$\pm$0.03     &0.4&
\\ SUBBOX 2c& 0.6$\pm$0.1     &0.4, 0.6&
\\ SUBBOX 2d& 0.1$\pm$0.2     &0.0, 0.4&\\ 
\hline
 SUBBOX 5a& 0.4$\pm$0.3     &0.2, 0.3& 0.49
\\ SUBBOX 5b& 0.4$\pm$0.02     &0.1, 0.7&
\\ SUBBOX 5c& 0.6$\pm$0.2     &0.6, 0.7 &
\\ SUBBOX 5d& 0.5$\pm$0.4     &0.6, 0.7&\\
\hline
 SUBBOX 6a& 0.5$\pm$0.1     &0.4& 0.50
\\ SUBBOX 6b& 0.0$\pm$0.1     &0.1, 0.6&
\\ SUBBOX 6c& -0.1$\pm$0.7     &0.6&
\\ SUBBOX 6d& 0.8$\pm$0.1     &0.6, 0.7&\\
\hline
 SUBBOX 9a& 0.6$\pm$0.2     & 0.6, 0.9& 0.45
\\ SUBBOX 9c&  0.0$\pm$4.1 &0.3, 0.4&
\\ SUBBOX 9d & -0.1$\pm$4.2       &0.2, 0.3&\\
\hline
 SUBBOX 10a& 0.1$\pm$0.2       &0.4&0.51
\\ SUBBOX 10b& 0.3$\pm$4.3   &0.5, 0.7&
\\ SUBBOX 10c& 0.3$\pm$0.1       &0.5, 0.7&
\\ SUBBOX 10d& 0.3$\pm$0.2       &0.3, 0.6&\\ 
\hline
 SUBBOX 13a& -0.9$\pm$4.6   &0.2, 0.5, 0.7& 0.33
\\ SUBBOX 13b& -0.5$\pm$0.7       &-0.1, 0.0, 0.4&
\\ SUBBOX 13c& -0.1$\pm$0.5       &0.2, 0.5&
\\ SUBBOX 13d& 0.1$\pm$0.6       &0.1, 0.7&
\\
\hline
\hline
\end{tabular}
\end{center}
\end{table}

Finally, we collect published integrated flux densities and errors for HB3 
at other frequencies in Table 4 and show them in Fig. 3. A new HB3 integrated flux
 density 34.7$\pm$12.9 at 2695 MHz is obtained by summing values of the integrated
 fluxes of the five subareas defined in Table 3. 
We have extrapolated total compact source fluxes 
to these other frequencies, using the 408-1420 MHz spectral index upper and lower 
limits. The compact sources contribution is  highest at 38 MHz at $\sim$ 9$\%$. 
It smoothly decreases to $\sim$ 4$\%$ at 3900 MHz. 
In the fitting below, we omit Trushkin et al. (1987)'s 
measurement of 70$\pm$5.0 Jy at 960 MHz. In fact, Trushkin himself (2002) predicted
that it should be about 48 Jy at 960 MHz in his spectrum calculation of HB3, i.e, 
about 30$\%$ of the measurement value is discarded. Considering Trushkin et al.'s 
observations at 3900 MHz and 3650 MHz have a large elliptical beam and suffer 
potentially from thermal contamination from the W3 HII complex, we also use 30$\%$ 
of their measurement as the error of each value. 
Since in all cases the
errors are dominated by systematic errors and are estimates by the authors without
details given it is difficult to interpret them. 
Here we take the published errors as 1-$\sigma$ errors but if they are
interpreted as 2-$\sigma$ errors, all of the $\chi^2$ values increase by a 
factor of 4.  
The best-fit least-squares fit for a single power-law has $\alpha$=0.54 and
$\chi^2$ of 5.3 for 11 degrees of freedom (dof).
Two power-laws results in significantly reduced $\chi^2$: 
from 38 MHz to 610 MHz, the best-fit $\alpha$ is 0.63 with  $\chi^2$ of 1.9 
for 6 dof, 
and from 408 MHz to 3900 MHz the best-fit is $\alpha$ is 0.32 with  
$\chi^2$ of 0.2 for 5 dof. 
Fig. 3 shows the three power-laws. Here we conclude that the fit with two 
power-laws better fits the data, but the significance of the improvement is 
difficult to
assess due to the difficulty in assessing the systematic errors in the data. 

\begin{figure}
\vspace{55mm}
\begin{picture}(0,80)
\put(-30,-20){\includegraphics{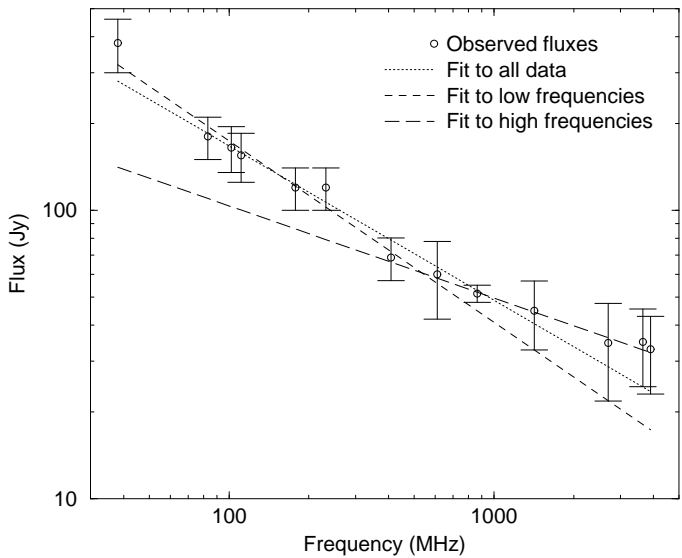}}
\end{picture}
\caption[xx]{Radio spectrum of HB3. 
The fit to all data has $\alpha$=0.54; the fit to  data between 38 MHz and 610 MHz 
has $\alpha$=0.63; and the fit to data between 408 MHz and 3900 MHz 
has $\alpha$=0.32}
\end{figure}

\begin{table*}
\begin{center}
\caption{Integrated flux and spectral index of the whole area and five subareas* 
of HB3 and of Compact Sources (CSs) at 1420 MHz and 408 MHz}
\setlength{\tabcolsep}{1mm}
\begin{tabular}{ccccccccc}
\hline
\hline
Freq.&CSs&HB3 Area&HB3+CSs&HB3 Sub-1 & HB3 Sub-2& HB3 Sub-3& HB3 Sub-4& HB3 Sub-5\\
\hline
MHz &Jy&Jy &Jy&Jy&Jy&Jy&Jy&Jy\\
\hline
\hline
408&5.3$\pm$0.4&68.6$\pm$11.5&73.9$\pm$11.9&41.4$\pm$7.9&12.5$\pm$0.9&4.1$\pm$0.1&13.6$\pm$0.9&1.3$\pm$0.2\\
1420&2.4$\pm$0.2&44.8$\pm$12.0&47.2$\pm$12.2&23.2$\pm$3.8&8.3$\pm$1.3&3.6$\pm$0.2&10.0$\pm$1.7&0.8$\pm$0.1\\
$\alpha$&0.64$\pm$0.10&0.34$\pm$0.15&0.36$\pm$0.15&0.47$\pm$0.16&0.33$\pm$0.07&0.11$\pm$0.01&0.34$\pm$0.06&0.34$\pm$0.08\\
\hline
\hline
\end{tabular}
\\**Subarea 1 is sum of B2, B3, B4, B6, B7, B8, B10, B11 and B12; Subarea 2 is sum of
 B1, B5 and B9; Subarea 3 is B13; Subarea 4 is sum of B14 and B15; Subarea 5 is B16.

\end{center}
\end{table*}

\begin{table}
\begin{center}
\caption{Published Integrated Flux Densities of HB3}
\setlength{\tabcolsep}{1mm}
\begin{tabular}{cccccc}
\hline
\hline
Freq. &Beamwidth&Flux Den. & references\\
MHz   &arcmin   & Jy               & \\
\hline
\hline
   38.0& 45$\times$45    &380.0 $\pm$80. & 1967, Caswell J.L. \\
  83.0 & 59$\times$31    &180.0 $\pm$30. & 1994, Kovalenko A.V. et al.\\
 102.0 & 48$\times$25    &165.0 $\pm$30. & 1994, Kovalenko A.V. et al.\\
 111.0 & 44$\times$23    &155.0 $\pm$30. & 1994, Kovalenko A.V. et al.\\
 178.0 & 23$\times$19    &120.0 $\pm$20. & 1967, Caswell J.L. \\
 232.0 &3.8$\times$4.3   &120.0 $\pm$20  & 1993, Zhu J. \\
 408.0 &3.5$\times$4.0   & 75.0 $\pm$15. & 1987, Landecker T.L. et al.\\
 408.0 &3.4$\times$3.8   & 68.6 $\pm$11.5& this paper\\
 610.0 &16 $\times$20    & 60.0 $\pm$18. & 1987, Landecker T.L. et al.\\
 863.0 &14.6$\times$14.3 & 51.5 $\pm$3.5 & 2003, Reich W. et al. \\
1420.4 &  1$\times$1.1   & 44.8 $\pm$12.0& this paper\\
 2695.0&  5$\times$5     & 34.0 $\pm$7.0 & 1974, Velusamy\&Kundu\\
2695.0 &3.4$\times$3.4   & 34.7 $\pm$12.9& this paper\\
3650.0 &  1$\times$10    & 35.0 $\pm$10.5& 1987, Trushkin S.A. \\
3900.0 &  1$\times$10    & 33.0 $\pm$9.9 & 1987, Trushkin S.A.  \\
\hline
\hline
\end{tabular}
\end{center}
\end{table}

\subsection{Comparison with Previously Published Integrated Fluxes}  
   Landecker et al. (1987) presented DRAO ST observations on HB3 at 408 MHz,
 and estimated its flux density 75$\pm$15Jy at 408 MHz after they considered that
the plateau of emission around W3 and the bright ridge near the apparent perimeter 
of HB3 are thermal and separated them from HB3. 
Landecker et al. considered the location border between W3 and HB3. 
Their work is confirmed and refined by our work based on the current
higher dynamic range data. 
We obtained maps including W3 and HB3 at 327 MHz from WENSS 
(resolution 54"x54"cosec($\delta$); Rengelink et al., 1997), 408 MHz and 1420 MHz 
from DRAO (resolution 
3.4'$\times$3.4'cosec($\delta$), and 1'$\times$1'cosec($\delta$), respectively), 
2695 MHz from the Effelsberg 100-m (resolution 4.3'; F\"urst et al., 1990) and 
4850 MHz from GB6 (resolution 3.7'$\times$3.3'; Condon et al., 1989). 
By comparing W3 and HB3 at different radio frequencies, a clear boundary
can be drawn. The thermal emission from W3 appears in similar regions at 
all frequencies and with no obvious decrease in strength, while the non-thermal 
emission from HB3 becomes progressively less prominent as the frequency increases. 
Because there is no apparent  
brightening of the SNR along the border with W3, 
this strongly supports Landecker et al.'s
 (1987) a conclusion that there is no interaction between HB3 and W3. 
Routledge et al.'s (1991) CO spectra observation reveals that HB3 has an interaction 
with molecular gas in which W3 is embedded, but no evidence for a direct interaction 
of W3 and HB3. Koralesky et al.'s (1998) maser study in the W3/HB3 complex 
also shows no evidence for any interaction between HB3 and W3.

Landecker et al.'s (1987) analysis on the uncertainty of the flux density
 of HB3  is consistent with our work, i. e., the uncertainty arises more from the 
difficulty of determining the background emission level than from imprecise 
separation of SNR from W3. 
We take two measures in order to calculate the 
integrated flux density of HB3. Firstly, we divide HB3 non-thermal region into 5 
subareas (see the division in Table 3). We remove all strong compact sources within 
HB3 and individually integrate all strong compact sources 
(integrated flux density is 5.3$\pm$0.4 Jy). 
After subtracting an average background emission value, we calculate an average pure 
emission value in each subarea, and then substitute the average value for the areas
removed due to compact sources. 
We then integrate five subareas and get the first estimate value of HB3 flux 
density (68.6$\pm$11.5 Jy). 
As a comparison procedure, we do the same removal of the compact sources, then we 
calculate a separate background value for each subarea and substitute the resulting
subarea average emission value for areas removed due to compact sources.  
Then we sum the integrated flux values for the five subareas to get the second
 estimated value of integrated flux density (72.9$\pm$9.9Jy).
Our results are consistent with each other and also with Landecker et al. 

Fesen et al. (1995) observed HB3 at both 408 MHz and 1420 MHz with the DRAO ST and 
estimated its integrated flux density (48.3$\pm$6.5 Jy at 408 MHz, 26.0$\pm$2.7 Jy 
at 1420 MHz), and calculated the spectrum index (0.5$\pm$0.2). 
The flux density they gave is rather low in comparison to previous authors' and 
ours. Their explanation that they subtracted the strong compact sources before 
integrating is not convincing because the 
total flux density of these compact sources we find is only 7$\%$ at 408 MHz 
and 5$\%$ at 1420 MHz of HB3's flux density. 
Fesen et al.'s (1995) 
calculation probably excluded too much area near the HB3-W3 boundary.  
 They obtained a spectral index value (0.64$\pm$0.01) 
using the T-T plot method, but there is no description of the selected area. 
 It is possible to get a steep spectral index 
from one area (e.g., subbox 2c or 6d) according to our results, but the average 
value from the whole area of HB3 is much less than 0.6.

Reich et al. (2003) executed their observations of HB3 at 35 cm wavelength with 
the Effelsberg 100-m radio telescope, and reported an integrated flux density 
of 51.5$\pm$3.5 Jy. There is no information concerning their estimation method 
except they mentioned the difficulty to obtain accurate flux density of HB3. 
Trushkin (2002) predicts about 57.8 Jy at 863 MHz based on 
all available data until 2002 and calculated a spectral index of 0.49. 

\subsection{Explanation of HB3 Spectral Index Variation}
  We have seen a gradient of thermal background emission from W3 decreasing toward 
the upper right in Fig. 1. By choice of background for the integrated flux
 measurement this smooth gradient is subtracted from the fluxes and spectral index. 
Similarly, by choosing small boxes for T-T plots, we make the T-T plot spectral 
index much less sensitive to a large scale background gradient.
An important point is that the spectral index of 0.32 obtained by fitting the higher 
frequency integrated flux data is consistent with the 408-1420 MHz 
integrated flux density based spectral index 
 0.34$\pm$0.15 and the best T-T plot spectral index 0.36. 

Now we discuss other properties of the spectral index variations, both
the variation with frequency (Fig. 3) and the variation of 408-1420 MHz spectral 
index with location within HB3 (Tables 1, 2 and 3). 
Since the integrated flux and T-T plot derived spectral indices agree, 
the filamentary emission (measured by the T-T plot method) 
and the total emission (filamentary
plus spatially smooth emission) have the same spectral index.
This agreement between T-T plot and integrated flux spectral index holds for
the whole of HB3 and also for the five subregions given in Table 3.
When we consider the T-T plot results in the finest detail (Table 2), we
find a range of spectral indices, generally in the range 0.1 to 0.7 
(with errors 0.1 - 0.2). 
The set of 408-1420 MHz spectral indices corresponds to a set of populations
of electrons with different energy indices.
Under the simplest assumption, each population has a power-law energy spectrum 
over a wide range of energies, so each population emits synchrotron radiation
over a wide range of radio frequencies with a single spectral index. 
Then the summed radio spectrum would have a larger contribution from the 
steeper index electron populations at low frequencies, and a larger contribution
from the flatter index electron populations at high frequencies.
This is consistent with what is observed (Fig. 3).
Alternatives, such as having electron populations with curved energy spectra in 
the normal sense with steeper energy index at higher energies, do not work
since they would produce a curved radio spectrum with steeper spectral index
at higher frequency, contrary to what is observed.

\section{Conclusion}
    Our 408 MHz and 1420 MHz maps of the SNR HB3 reveal similar but more detailed
weak emission structure features than previously published maps. 
Considering a background emission gradient across HB3 area, we
give HB3's flux density at 408 MHz, 1420 MHz and 2695 MHz. The
integrated flux-density based spectral index between 408 MHz and 1420 MHz is 
0.34$\pm$0.15. The averaged T-T plot spectral index is 0.36.  Based on the 
analysis of our data and previous data, we reconcile our 
spectral index result with the previous authors' 0.6. 
 A viable explanation is that on small spatial scales the spectral
index varies within the range 0.1-0.7. Thus the low frequency data mainly reflect 
the steeper emission components and the high frequency data mainly reflects 
the flatter emission components.

\begin{acknowledgements}
We thank the referee for his constructive suggestions which helped to improve 
the paper. Dr. R. Kothes at the DRAO provided information on calibration errors 
of the CGPS data.  W.W. Tian thanks 
the MoST for grant NKBRSF 2003CB716703 and NSF grant in China. We acknowledge 
support from the Natural Sciences and Engineering Research Council of Canada. 
The DRAO is operated as a national facility by the National Research Council of 
Canada. The Canadian Galactic Plane Survey is a Canadian project with international 
partners. 
\end{acknowledgements}

\end{document}